\documentstyle[12pt,epsfig]{article}
\def\beq{\begin{equation}}
\def\eeq{\end{equation}}
\def\be{\begin{equation}}
\def\ee{\end{equation}}
\def\bea{\begin{eqnarray}}
\def\eea{\end{eqnarray}}
\def\nnb{\nonumber}

\newcommand{\wti}{\widetilde}
\newcommand{\gsim}{\lower.7ex\hbox{$\;\stackrel{\textstyle>}{\sim}\;$}}
\newcommand{\lsim}{\lower.7ex\hbox{$\;\stackrel{\textstyle<}{\sim}\;$}}

\begin{document}

\begin{center}
\vspace{-3ex}{
                      \hfill hep-th/0605016}\\[2mm]
{\LARGE\bf  
WZW action in odd dimensional gauge theories}

\vspace{0.6cm}
{\bf Wei Liao }

\vspace{0.3cm}
{\it TRIUMF, 4004 Wesbrook Mall, Vancouver, V6T 2A3, Canada}
\end{center}
\begin{abstract}
It is shown that Wess-Zumino-Witten (WZW) type actions can be
constructed in odd dimensional space-times using Wilson
line or Wilson loop. WZW action constructed using Wilson
line gives anomalous gauge variations and the WZW action
constructed using Wilson loop gives anomalous chiral
transformation. We show that pure gauge theory including
Yang-Mills action, Chern-Simons action and the WZW action
can be defined in odd dimensional space-times with even
dimensional boundaries. Examples in 3D and 5D are given.
We emphasize that this offers a way to generalize gauge
theory in odd dimensions. The WZW action constructed
using Wilson line can not be considered as action localized
on boundary space-times since it can give anomalous gauge
transformations on separated boundaries. We try to show that
such WZW action can be obtained in the effective theory 
when making localized chiral fermions decouple.

\end{abstract}

\section{Introduction}\label{sec1}
Wess-Zumino-Witten action~\cite{wznw} as originally
constructed for low energy mesons is known to produce
the prediction of chiral anomaly at the level of Nambu-Goldstone boson. It is
constructed in the non-linear sigma model using the field 
$\Sigma=e^{2i \pi/F_\pi}$
where $\pi=\sum_a T^a \pi^a$ is the meson field. Under
a left-right transformation $\Sigma'= U_L \Sigma U^{-1}_R$
the action is not invariant and gives anomalous
transformation. 
WZW action achieved beautiful success in its
electromagnetic version which fixes the strength of
processes like $\pi^0 \to 2 \gamma$ and $K^+ K^- \to 3 \pi$.

WZW action is of broader interests in quantum field theories.
The availability of such action offers an alternative way 
to construct anomaly-free 
gauge theories. The canonical way to make the chiral gauge theory
anomaly-free is to arrange the fermion content in such a way that
anomaly contributions of individual fermions cancel in the
sum. This is exactly what happened in the Standard Model.
The alternative way using the WZW action states
that one can take fermion content which has non-vanishing
gauge anomaly. The gauge anomalies in the fermionic sector
and WZW part can be arranged to cancel.

The canonical way and the alternative to build anomaly-free
theory can be connected by studying the decoupling limit
of heavy fermions in a gauge theory with anomaly-free
fermionic content. Heavy fermions in the theory which decouple 
can be arranged to give non-zero gauge anomalies.
The consistency of the
gauge theory in the effective theory is guaranteed 
with the appearance of the WZW action.
For example some fermions, say $\Psi$'s, get massive from
the following term.
\bea
\lambda ({\bar \Psi}_L \Phi \Psi_R + h.c.) \nnb
\eea
As long as $\lambda$ is large enough one can integrate
out the heavy degrees of freedom $\Psi$'s. The effective
action will be the action of the field $\Phi$, the
gauge field and the light fermions $\psi$'s. 
The effective action of $\Phi$
and the gauge field has to reproduce the gauge anomaly
of $\Psi$'s and cancels the gauge anomaly contributed by
light fermions $\psi$'s.
So the effective theory is still anomaly-free.

Gauge anomalies and chiral anomalies studied in literature
can be classified into anomalies of LA form and VA form.
Consider a theory with Lagrangian
\bea
{\cal L} &&={\bar \psi} i \gamma^\mu (\partial_\mu - i V_\mu
-i \gamma_5 A_\mu) \psi, \nnb \\
&& ={\bar \psi}_L i\gamma^\mu (\partial_\mu-i A_{L\mu}) \psi_L
+{\bar \psi}_R i\gamma^\mu (\partial_\mu-i A_{R\mu}) \psi_R, \nnb
\eea
where 
\bea
V_\mu= \frac{1}{2}(A_{R\mu}+A_{L\mu}),~~
A_\mu= \frac{1}{2}(A_{R\mu}-A_{L\mu}),~~
\psi_{L,R}=\frac{1\mp \gamma_5}{2}\psi. \nnb
\eea
$V_\mu=\sum_a T^a V^a_\mu$ is the vector field and 
$A_\mu=\sum_a T^a A^a_\mu$ is the axial-vector field.
The coupling constants are absorbed into $V_\mu$ and $A_\mu$.
$T^a$ is the generator of the gauge group $G$.
$V_\mu$ and $A_\mu$ which can be gauge fields or auxiliary
fields are taken as external fields in computing one-loop
anomalies.
The LR form of the anomaly is
\bea
D^\mu J^a_{L\mu} &&= G^a_L(A_L)
= \frac{1}{24 \pi^2} \varepsilon^{\mu \nu \rho \sigma}
Tr[T^a \partial_\mu (A_{L\nu} \partial_\rho A_{L\sigma}
-\frac{i}{2} A_{L\nu}A_{L\rho}A_{L\sigma})], \label{ano1} \\
D^\mu J^a_{R\mu} &&= G^a_R(A_R)
= -\frac{1}{24 \pi^2} \varepsilon^{\mu \nu \rho \sigma}
Tr[T^a \partial_\mu (A_{R\nu} \partial_\rho A_{R\sigma}
-\frac{i}{2} A_{R\nu}A_{R\rho}A_{R\sigma})], \label{ano2}
\eea
where $\varepsilon^{\mu \nu \rho \sigma}$ is the
anti-symmetric tensor with $\varepsilon^{0123}=1$, and
\bea
&&J^a_{L\mu}={\bar \psi}_L T^a \gamma_\mu \psi_L, ~~
J^a_{R\mu}={\bar \psi}_R T^a \gamma_\mu \psi_R. \label{current1}
\eea
(\ref{ano1}) and (\ref{ano2}) are called LR form because in 
computing it the loop momentum
is labeled in such a way that the anomaly takes the left-right symmetric
form. Gauge anomaly of this form is also called consistent gauge
anomaly. It is known that there is no unique way to label the
loop momentum and anomaly can be shifted between vector current
and axial-vector current~\cite{wein}. The loop momentum can 
also be labeled in such
a way that the vector current is covariantly conserved and 
then anomaly is completely shifted to the axial-vector current, that
is the VA form of the anomaly~\cite{Bard}
\bea
D^\mu J^{5a}_\mu && = G^a_A(V,A) \nnb \\
&&=-\frac{1}{4 \pi^2} \varepsilon^{\mu \nu \rho \sigma}
Tr\{ T^a [\frac{1}{4} V_{\mu \nu} V_{\rho \sigma}
+\frac{1}{12} A_{\mu \nu} A_{\rho \sigma} \nnb \\
&&+ \frac{2 i}{3}(A_\mu A_\nu V_{\rho \sigma}+A_\mu V_{\nu \rho} A_\sigma
+V_{\mu \nu} A_\rho A_\sigma) -\frac{8}{3} A_\mu A_\nu A_\rho A_\sigma] \},
\label{ano3}
\eea
where
\bea
&& J^{5a}_\mu ={\bar \psi} T^a \gamma_\mu \gamma_5 \psi,
\label{current2} \\
&&V_{\mu \nu}=\partial_\mu V_\nu -\partial_\nu V_\mu -i [V_\mu,V_\nu]
-i[A_\mu,A_\nu], \label{streng1} \\
&&A_{\mu \nu}=\partial_\mu A_\nu -\partial_\nu A_\mu -i [V_\mu,A_\nu]
-i [A_\mu,V_\nu]. \label{streng2}
\eea
In a theory with both left and right gauged symmetries
it is natural to take the LR form of the anomaly. On the
other hand, if only the vector part of the symmetry is
gauged it is more natural to shift the anomaly to be
in the axial-vector current and take the vector current
conserved. This form of anomaly when applied to
QED coincides with the original chiral anomaly obtained in 
~\cite{ABJ}.
The WZW action which produces anomalies at the level of
Nambu-Goldstone boson can also be built to produce anomalies of
LR form or VA form ~\cite{wein}.

In 5D models it is interesting to notice that a Wilson line
along the extra space-like dimension
\bea
W(x^\mu)= {\cal P} e^{i\int^{\pi R}_0 dy ~A_4(x^\mu,y)} , \nnb
\eea
transforms as bifundamntal, {\it i.e.} $W'=U(y=0) W U^{-1}(y=\pi R)$.
$A_4$ is the gauge field of the fouth space-like dimension.
Since $U(y=0)$ and $U(y=\pi R)$ are gauge transformations
at different points in the extra dimension, they can be
considered as independent gauge transformations from 4D point
of view. Then $W$ is similar to the
$\Sigma=e^{2 i \pi/F_\pi}$ field in the sigma model.
As will be seen in the following we can build 
WZW type action using the Wilson line or the Wilson loop. The WZW action
constructed using Wilson line give gauge anomalies (of LR form).
The WZW action constructed using Wilson
loop is gauge invariant and gives anomalous variation
under chiral transformation (of VA form). 
It will be shown that it is possible to construct pure
gauge theories with the WZW action which gives anomalous
gauge transformation. The theory is defined on odd dimensional
space-times (3D and 5D) and it consists of the pure Yang-Mills,
Chern-Simons and the WZW action constructed using the
Wilson line. The theory is made gauge invariant by requiring
the anomalous gauge variations of the Chern-Simons action and
the WZW action have the same magnitude and the opposite sign
and hence cancel on the boundary space-times.

We try to show that such kind pure gauge theory can arise
as an effective theory of a gauge invariant theory with
localized fermions. Chiral fermions $\Psi_{L,R}$ charged
under gauge group are localized on different boundary branes.
Hence the gauge anomalies are localized on the branes.
The theory is made gauge invariant by including the
Chern-Simons term in the bulk and requiring that its
anomalous gauge variations cancel those localized on the
boundary space-times. Chiral fermions localized on boundary 
space-times can couple
to the Wilson line $W$ which links the two boundaries with
interaction
\bea
m({\bar \Psi}_L W \Psi_R +h.c.) \nnb
\eea
One can integrate out the fermion $\Psi_{L,R}$ by sending
$m \to \infty$. As is required by the
consistency of the theory, the effective theory after
integrating out fermion should also be anomaly-free. This is
obtained by the appearance of the WZW action built of
$W$ in the effective theory.

We motivate that we can generalize the gauge theory in
odd dimensional space-time using the WZW action constructed
using Wilson line. The Wilson line used in the theory
links the boundary branes. The WZW action constructed
like this gives non-local interaction in odd dimensional
space-time and can not be considered as action localized
on even dimensional boundaries.
The present work is inspired by a series of recent works
~\cite{Hill1,Hill2} by C. Hill who constructed 5D models
with localized chiral fermions, hence localized gauge anomalies,
on separate branes.  In section \ref{sec2} we illustrate the point
of our paper with a 3D example. We construct the WZW action
in 3D space-time with boundaries using Wilson line and show 
that it gives anomalous
gauge transformations on boundaries which are gauge anomalies
of the consistent form (LR form). We construct a pure gauge theory in 3D
which includes Yang-Mills action, Chern-Simons and WZW action.
In section \ref{sec3} we give
example in 5D which is a bit more complicated. We study
how such WZW action arises from integrating out heavy chiral fermions 
localized on different branes.
In section \ref{sec4} we construct WZW action using Wilson
loop. The action constructed is gauge invariant but gives anomalous chiral 
transformation (of VA form).
We also try to obtain this action from decoupling
heavy fermions localized on boundaries.
We comment and summarize in section \ref{sec5}.

\section{WZW action in 3D gauge theory}\label{sec2}
We begin with the simple example in 3 dimension. 
Consider 3D flat spacetime $\Sigma_3=M_2 \times [0,\pi R]$ with coordinates
$x^M=(x^\mu,x^2=y)$ ($\mu=0,1$). ${\cal M}_2$ is two dimensional
Minkowski space-time. There are two 2D boundary branes L and R
at $y=0$ and $y=\pi R$ separately. Gauge field $A_M$ of gauge group
$G$ propagates in the 3D space-time. Gauge group on boundaries
$G_L$ and $G_R$ are determined by the boundary conditions and
can be smaller than $G$. For simplicity we assume $G_L=G_R=G$.
We introduce gauge fields on the boundaries which are
obtained from reducing $A_M$ to the boundaries:
\bea
A_{L\mu}=A_\mu(x^\mu,y=0), ~~ A_{R\mu}=A_\mu(x^\mu,y=\pi R).
\label{def15}
\eea
$A_M=\sum_a T^a A^a_M$, similarly for $A_{L,R\mu}$.
$A_M$ and $A_{L,R\mu}$ are defined as having dimension $[M]$.
We then introduce $U_L$ and $U_R$,
\bea
U_L(x^\mu)=U(x^\mu,y=0), ~~~U_R(x^\mu)=U(x^\mu,y=\pi R).
\label{def16}
\eea
So the gauge transformation
\bea
A'_M(x^\mu,y) = U(x^\mu,y) A_M(x^\mu,y) U^{-1}(x^\mu,y)
+i U(x^\mu,y) \partial_M  U^{-1}(x^\mu,y),
\label{gaugtr1}
\eea
when reduced to L and R boundary branes are written as
\bea
A'_{L \mu}&&=U_L(x^\mu) A_{L \mu} U^{-1}_L(x^\mu)
+i U_L(x^\mu)\partial_\mu  U^{-1}_L(x^\mu), 
\label{gaugtr2} \\
A'_{R \mu}&&=U_R(x^\mu) A_{R \mu} U^{-1}_R(x^\mu)
+i U_R(x^\mu)\partial_\mu  U^{-1}_R(x^\mu).
\label{gaugtr2a}
\eea
A Wilson line linking two boundaries is defined as
\bea
W(x^\mu)= {\cal P} e^{i\int^{\pi R}_0 dy ~A_2(x^\mu,y)} ,
\label{WilL2}
\eea
where ${\cal P}$ is the path-ordering operator and
$A_2$ is the gauge field along the compact space-like
dimension. Under the gauge transformation (\ref{gaugtr1}),
the Wilson line transforms as
\bea
W'(x^\mu)&&= U(x^\mu,0) W(x^\mu) U^{-1}(x^\mu,\pi R) \nnb \\
&& =U_L(x^\mu) W(x^\mu) U^{-1}_R(x^\mu).
\label{gaugtr3}
\eea
We also introduce $W_y(x^\mu,y)$
\bea
W_y(x^\mu,y)= {\cal P} e^{i\int^y_0 dy' ~A_2(x^\mu,y')}.
\label{WilL2a}
\eea
$W_y(x^\mu,y)$ satisfies
\bea
W_y(x^\mu,y=0)=1, ~~ W_y(x^\mu,y=\pi R)=W(x^\mu). 
\label{intercon1}
\eea
For gauge transformed $W'$ one can also introduce $W'_y$.
Condition (\ref{intercon1}) is also satisfied for gauge 
transformed $W'_y$. We have 
\bea
W'_y(x^\mu,y)=U(x^\mu,0) W_y(x^\mu,y) U^{-1}(x^\mu,y).
\label{gaugtr3a}
\eea
We note that $W_y(x^\mu,y=0)=1$ and the configuration of $W_y$ can 
be taken as a mapping of ${\wti \Sigma}_3$ to the space of gauge group $G$.
${\wti \Sigma}_3$ is $\Sigma_3$ with the boundary
at $y=0$ shrinking to a point. So ${\wti \Sigma}_3$ has a single boundary
at $y=\pi R$:
$\partial {\wti \Sigma}_3={\cal M}_2$.

An anomalous action can be constructed as
\bea
\Gamma_{WZW}&&=\frac{1}{12\pi} \int_{\Sigma_3} d^3x ~\varepsilon^{RST}
~Tr[(\partial_R W_y) W^{-1}_y (\partial_S W_y) W^{-1}_y 
(\partial_T W_y) W^{-1}_y ] \nnb \\
&&+\frac{i}{4\pi}\int_{{\cal M}_2} d^2x ~ \varepsilon^{\mu\nu}
~Tr[A_{L\mu}W_{L\nu}+A_{R\mu}W_{R\nu}
-i A_{R\mu}W^{-1} A_{L\nu} W],
\label{act7}
\eea
where $R,S,T$ run over $0,1,2$, $\varepsilon^{012}=\varepsilon^{01}=1$
and 
\bea
W_{L\mu}= (\partial_\mu W)W^{-1}, ~~W_{R\mu}=W^{-1}(\partial_\mu W).
\label{def17}
\eea

(\ref{act7}) is formally of the 2D WZW action.
However its interpretation and physical
content is quite different. $W_y$ is not an auxiliary
extension of $W$ to the third auxiliary dimension and the
3D integration is not in an auxiliary space-time either. Furthermore
this action defined using Wilson line gives non-local interaction
for $A_L$ and $A_R$ at different branes
and can not be interpreted as action localized on boundaries.
One can see that last term in (\ref{act7}) mixes gauge fields
on two boundaries $A_L$ and $A_R$ via the link field $W$.
One can also make the point clear by studying the gauge
transformation properties of action (\ref{act7}).
Using (\ref{gaugtr1}), (\ref{gaugtr3a}), (\ref{gaugtr2})
and (\ref{gaugtr2a}), action (\ref{act7}) transforms under
an infinitesimal transformation as
\bea
\delta \Gamma_{WZW}=-\frac{1}{4\pi}\int_{{\cal M}_2}
\varepsilon^{\mu\nu} ~Tr[\epsilon_L ~\partial_\mu A_{L\nu}]
+\frac{1}{4\pi}\int_{{\cal M}_2}
\varepsilon^{\mu\nu} ~Tr[\epsilon_R ~\partial_\mu A_{R\nu}],
\label{gaugtr4}
\eea
where $\epsilon_{L,R}$ is given by $U=e^{i \epsilon}$ 
which approaches unity at infinity and
\bea
U_{L,R}=e^{i \epsilon_{L,R}}, ~~\epsilon_L(x^\mu)=\epsilon(x^\mu,y=0),
~~\epsilon_R(x^\mu)=\epsilon(x^\mu,y=\pi R),
\label{def18}
\eea
(\ref{gaugtr4}) is of the form of the consistent gauge anomaly 
(of LR form) in two
dimension~\cite{BZ}. We note that $A_L$ and $A_R$ are gauge fields
on two boundaries. Under gauge transformation the action
gives anomalous variations on two boundaries. This can not
be achieved by a single WZW action localized on boundary.
\\

\begin{figure}%[t]
\begin{center}
\psfig{figure=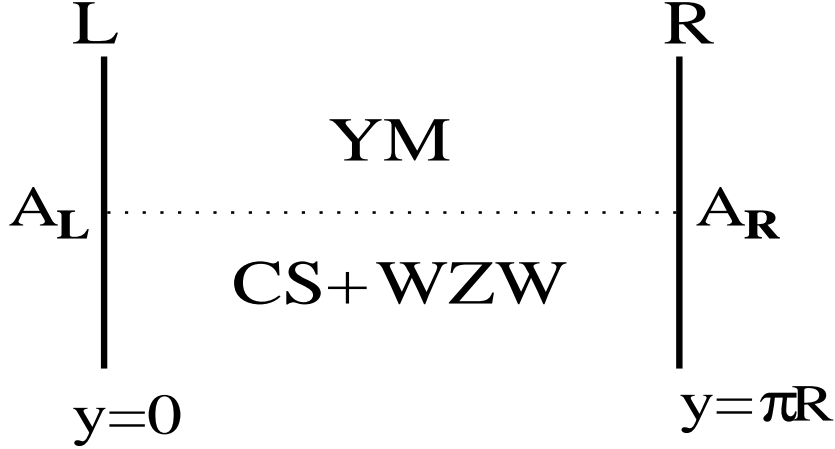,height=6cm,width=11cm}
\vskip 0.5cm
\caption{\small Pure gauge theory in odd dimensional (3D or 5D) space-time.
Gauge fields on L and R branes are induced by gauge field in bulk:
$A_{L\mu}(x^\mu)=A_\mu(x^\mu,y=0)$ and $A_{R\mu}(x^\mu)=A_\mu(x^\mu,y=\pi R)$.}
\label{gauge}
\end{center}
\end{figure}

A pure gauge theory can be defined in $\Sigma_3$ using (\ref{act7})
together with Chern-Simons action, as shown in Fig. \ref{gauge}.
The action of the theory is
\bea
\Gamma= \Gamma_{YM}+\Gamma_{CS}+\Gamma_{WZW}.
\label{def19}
\eea
$\Gamma_{YM}$ is the Yang-Mills kinetic action which is
itself gauge invariant and will not elaborated in the following.
The Chern-Simons action is
\bea
\Gamma_{CS}= -\frac{1}{4\pi} \int_{\Sigma_3} ~d^3x ~\varepsilon^{RST}
~Tr[A_R \partial_S A_T-\frac{2i}{3} A_R A_S A_T].
\label{def20}
\eea
Under infinitesimal gauge transformation given in (\ref{def18}),
(\ref{gaugtr1}), (\ref{gaugtr2}) and (\ref{gaugtr2a}), action
$\Gamma_{CS}$ transforms as
\bea
\delta \Gamma_{CS}&&= -\frac{1}{4\pi} \int_{\Sigma_3} 
d^3x ~\varepsilon^{RST} \partial_R Tr[\epsilon ~\partial_S A_T] \nnb \\
&&=-\frac{1}{4\pi} \int_{{\cal M}_2} d^2x ~\varepsilon^{\mu\nu}
Tr[\epsilon_R ~\partial_\mu A_{R\nu}]
+\frac{1}{4\pi} \int_{{\cal M}_2} d^2x ~\varepsilon^{\mu\nu}
Tr[\epsilon_L ~\partial_\mu A_{L\nu}],
\label{gaugtr5}
\eea
where the integration over total divergence in $\Sigma_3$
is reduced to the integration on boundaries in ${\cal M}_2$.
It is clear that (\ref{gaugtr5}) and (\ref{gaugtr4}) cancel
in (\ref{def19}), {\it i.e.} gauge invariance is achieved
in (\ref{def19}): $\delta \Gamma=0$.
(\ref{def19}) defines a generalized gauge theory in 3D space-time
with boundaries.
In this theory gauge anomalies in Chern-Simons part and 
the WZW part are canceled on the boundary space-times.

\section{WZW action in 5D gauge theory}\label{sec3}
In this section we present a 5D example which is a bit more complicated.
Consider a 5D flat space-time $\Sigma_5={\cal M}_4\times [0,\pi R]$
with coordinate $x^M=(x^\mu,x^4=y)$ ($\mu=0,1,2,3$). 
${\cal M}_4$ is 4 dimensional Minkowski space-time.
Gauge fields of gauge group $G$ propagate in the bulk. There are
two boundary branes in $\Sigma_5$, namely brane L at $y=0$ 
and brane R at $y=\pi R$. The gauge groups $G_L$ and $G_R$
on the boundary branes L and R are determined
by the boundary conditions and can be smaller than the
gauge group $G$ in the bulk. For simplicity we assume $G_L=G_R=G$.
We introduce gauge fields on the boundary branes L and R as
$A_{L\mu}$ and $A_{R\mu}$, defined using Eq. (\ref{def15})
with $\mu=0,1,2,3$. Eq. (\ref{def16}), (\ref{gaugtr1}), (\ref{gaugtr2})
and (\ref{gaugtr2a}) give the gauge transformations of $A_M$ and
$A_{L,R\mu}$ with $M=0,1,2,3,4$ and $\mu=0,1,2,3$.

We also introduce Wilson line $W(x^\mu)$ and $W_y(x^\mu,y)$
\bea
W(x^\mu)= {\cal P} e^{i\int^{\pi R}_0 dy ~A_4(x^\mu,y)} ,
~~W_y(x^\mu,y)= {\cal P} e^{i\int^y_0 dy' ~A_4(x^\mu,y')} ,
\label{WilL}
\eea
where ${\cal P}$ is the path-ordering operator.
Under gauge transformation (\ref{gaugtr1}) they transform as
\bea
W'(x^\mu)
&& =U_L(x^\mu) W(x^\mu) U^{-1}_R(x^\mu),
\label{gaugt3} \\
W'_y(x^\mu,y)&&= U(x^\mu,0) W_y(x^\mu,y) U^{-1}(x^\mu,y).
\label{gaugt3a}
\eea

$W_y$ satisfies the condition
\bea
W_y(x^\mu,y=0)=1,~~~W_y(x^\mu,y=\pi R)=W(x^\mu).
\label{def7}
\eea
Since $W_y(x^\mu,y=0)=1$ the configuration of $W_y(x^\mu,y)$
can be considered as a mapping of ${\wti \Sigma}_5$ to space of the
gauge group $G$ where ${\wti \Sigma}_5$ is $\Sigma_5$ with
the boundary at $y=0$ shrinking to a point. So ${\wti \Sigma}_5$
has a single boundary: ${\cal M}_4 = \partial {\wti \Sigma}_5$.

\subsection{A pure gauge theory with WZW action}\label{sec31}
WZW action is defined as
\bea
\Gamma_{WZW}&&= \frac{-i}{240 \pi^2} \int_{\Sigma_5} d^5x ~Tr[
\varepsilon^{MNRST}\frac{\partial W_y}{\partial x^M} W^{-1}_y
\frac{\partial W_y}{\partial x^N} W^{-1}_y
\frac{\partial W_y}{\partial x^R} W^{-1}_y
\frac{\partial W_y}{\partial x^S} W^{-1}_y
\frac{\partial W_y}{\partial x^T} W^{-1}_y]
\nnb \\
&& +\frac{-i}{48\pi^2} \int_{{\cal M}_4} d^4 x
~\varepsilon^{\mu\nu\rho\sigma} ~ Tr[ 
~\frac{1}{2}(W_{R_\mu}A_{R\nu}W_{R\rho}A_{R\sigma}
-W_{L_\mu}A_{L\nu}W_{L\rho}A_{L\sigma})  \nnb \\
&&
+W_{L\mu}(A_{L\nu}\partial_\rho A_{L\sigma}
+(\partial_\nu A_{L\rho})A_{L\sigma}-i A_{L\nu} A_{L\rho} A_{L\sigma}
-iW_{L\nu}W_{L\rho}A_{L\sigma})
+ (L \to R) ~] \nnb \\
&&+\frac{1}{48\pi^2} \int_{{\cal M}_4} d^4 x
~\varepsilon^{\mu\nu\rho\sigma} ~ Tr[
i A_{L\mu}W A_{R\nu}W^{-1} W_{L\rho}W_{L\sigma}
-i A_{R\mu}W^{-1} A_{L\nu}W W_{R\rho}W_{R\sigma} \nnb \\
&&+(A_{R\mu}\partial_\nu A_{R\rho}
+(\partial_\mu A_{R\nu})A_{R\rho}-i A_{R\mu} A_{R\nu} A_{R\rho})
W^{-1} A_{L\sigma} W- (L \leftrightarrow R, W^{-1} \leftrightarrow W) \nnb \\
&& +A_{R\mu}W^{-1}A_{L\nu}W A_{R\rho} W_{R\sigma}
+ A_{L\mu}W A_{R\nu}W^{-1} A_{L\rho} W_{L\sigma}
-\frac{i}{2} A_{R\mu} W^{-1} A_{L\nu} W A_{R\rho} W^{-1} A_{L\sigma} W \nnb \\
&&+i (\partial_\mu A_{R\nu}) (\partial_\rho W^{-1}) A_{L\sigma} W
- (L \leftrightarrow R, W^{-1} \leftrightarrow W) . ~]
\label{WZW1}
\eea
where $M,N,R,S,T$ run over $0,1,2,3,4$,
$\varepsilon^{01234}=\varepsilon^{0123}=1$
($\varepsilon_{01234}=-\varepsilon_{0123}=1$) and
\bea
W_{L\mu} =\frac{\partial W}{\partial x^\mu} W^{-1}, ~~~
W_{R\mu} =W^{-1} \frac{\partial W}{\partial x^\mu}.
\label{def6}
\eea
$M,N,R,S,T$ run over $0,1,2,3,4$.
(\ref{WZW1}) is of the form 
of the 4D WZW action~\cite{cgws,krs,kt,mm,manes}.
The interpretation and the physics content
are however quite different. (\ref{WZW1})
defines non-local interactions of gauge fields
on two boundaries, $A_L$ and $A_R$, via the link field $W$
and can not be understood as action localized on the boundaries. 
Further discussions on this action closely follow
the discussions on (\ref{act7}).

Under infinitesimal transformation $U(x^\mu,y)$ which
approaches unity at infinity,
\bea
&U=e^{i \epsilon}, ~~U_L=U(y=0)=e^{i \epsilon_L}, 
~~U_R=U(y=\pi R)=e^{i \epsilon_R},
\label{gaugt3b} \\
&\epsilon_L(x^\mu)=\epsilon(x^\mu,y=0),~~
\epsilon_R(x^\mu)=\epsilon(x^\mu,y=\pi R),
\label{gaugt3c}
\eea
we obtain
\bea
\delta \Gamma_{WZW}=
- \frac{1}{24\pi^2}\int_{{\cal M}_4}d^4x ~\omega_4^1(A_L,\epsilon_L)
+\frac{1}{24\pi^2} \int_{{\cal M}_4}d^4x ~\omega_4^1(A_R,\epsilon_R),
\label{gaugt5}
\eea
where $\omega_4^1$ is
\bea
\omega_4^1(B_\mu(x^\mu),\varepsilon(x^\mu))
=Tr[\varepsilon^{\mu\nu\rho\sigma} 
\varepsilon(x^\mu) \partial_\mu
(B_\nu \partial_\rho B_\sigma
-\frac{i}{2}B_\nu B_\rho B_\sigma)].
\label{def5}
\eea
(\ref{gaugt5}) takes the form of the 4D gauge anomaly (LR form)
~\cite{cgws,krs,kt,mm}. 
Its interpretation is 
that under the gauge transformation (\ref{gaugtr1}) and (\ref{gaugt3a})
in the bulk the action (\ref{WZW1}) gives anomalous gauge 
transformations on two separated boundaries. This can not
be achieved by a single WZW action localized on boundaries.
\\

A pure gauge theory can be defined on $\Sigma_5$ using
(\ref{WZW1}) and the Chern-Simons action, as illustrated in Fig. \ref{gauge}.
The action of the theory is
\bea
\Gamma= \Gamma_{YM} +\Gamma_{CS}+\Gamma_{WZW}.
\label{def4}
\eea
The pure Yang-Mills action $\Gamma_{YM}$ is itself gauge invariant.
The Chern-Simons action to be included is given as
\bea
\Gamma_{CS}(A_M)=
\frac{-1}{24\pi^2}~ \int_{\Sigma_5} d^5 x ~\omega_5(A_M,F_{MN}),
\label{C-S1}
\eea
where
\bea
\omega_5( A_M,F_{MN})= \varepsilon^{MNRST}
Tr[\frac{1}{4} A_M F_{NR}F_{ST}
+\frac{i}{4} A_M A_N A_R F_{ST}-\frac{1}{10} A_M A_N A_R A_S A_T].
\label{C-S2}
\eea

It can be checked that
under infinitesimal transformation (\ref{gaugt3b}) and (\ref{gaugt3c}),
action (\ref{C-S1}) gives a total derivative in the integrand and the 
integration $\Sigma_5$ is then reduced to the boundary ~\cite{zwz}:
\bea
\delta \Gamma_{CS} && = \frac{1}{48\pi^2}~ \int_{\Sigma_5} d^5 x
~\varepsilon^{MNRST} \partial_M Tr[ (\partial_N \epsilon)
(A_R \partial_S A_T+(\partial_R A_S) A_T -i
A_R A_S A_T)] \nnb \\
&& =-\frac{1}{24\pi^2}~ \int_{{\cal M}_4} d^4 x 
~\omega_4^1(A_{R\mu},\epsilon_R) 
+ \frac{1}{24\pi^2}~ \int_{{\cal M}_4} d^4 x
~\omega_4^1(A_{L\mu},\epsilon_L),
\label{gaugt4}
\eea
where 
$\varepsilon^{4\mu\nu\rho\sigma}
=\varepsilon^{\mu\nu\rho\sigma}$ and integration by
part on ${\cal M}_4$ has been used.

Using (\ref{gaugt5}) and (\ref{gaugt4}) one can see
$\delta \Gamma=0$. The gauge invariance of the theory
is achieved by making the anomalous gauge variations of
Chern-Simons part and the WZW part cancel on the boundary
space-times.

\subsection{WZW action from decoupling fermion}\label{sec32}
In this subsection we study a gauge theory with chiral fermions
localized on boundaries and Chern-Simons action in the 5D bulk.
We try to obtain the WZW action described in the last
subsection in the effective theory when making the
localized chiral fermions heavy and decouple.
We work again in $\Sigma_5$.
We introduce $\psi_L$ on brane L and
$\psi_R$ on brane R. They are charged under the gauge
group $G_L$ and $G_R$, namely coupled to $A_L$ and $A_R$ separately.
The gauge invariance of the theory is achieved by making
the anomalous gauge variations of bulk part, {\it i.e.}
of Chern-Simons part at the classical level, cancel the 
anomalous gauge variations given by the boundary fermions
which are at the quantum level. 

This cancellation is possible by noting that
the chiral gauge theories on the 4D boundaries are known to
be anomalous, {\it i.e.} the currents are not
covariantly conserved on two boundaries.
The non-conservation is understood at the functional level as arising
from non-invariance of the fermionic measure under the left-right 
transformation~\cite{Fuji,bert}.
Under infinitesimal transformation (\ref{gaugt3b}) and (\ref{gaugt3c})
\bea
\psi'_L=U_L \psi_L, ~~\psi'_R=U_R \psi_R,
\label{gaugt3d}
\eea
where $U_{L,R}$ is defined in (\ref{def16}), the functional measure
changes
\bea
D\psi'_L D{\bar \psi}'_L=D\psi_L D{\bar \psi}_L
e^{-i \int d^4 x~ \epsilon^a_L G^a_L(A_L)},~~
D\psi'_R D{\bar \psi}'_R=D\psi_R D{\bar \psi}_R
e^{-i \int d^4 x~ \epsilon^a_R G^a_R(A_R)},
\label{gaugt6}
\eea
where $\epsilon_{L,R}=\sum_a T^a \epsilon^a_{L,R}$.
$G_L(A_L)$ and $G_R(A_R)$ are given in (\ref{ano1})
and (\ref{ano2}). 

We consider the action
\bea
\Gamma= \Gamma_{YM}+\Gamma_{CS} +\Gamma_{eff},
\label{def8}
\eea
$\Gamma_{YM}$ is the pure Yang-Mills action which is gauge
invariant itself and will not be elaborated in the following.
$\Gamma_{CS}$ is the Chern-Simons action given in (\ref{C-S1}).
It is not gauge invariant and gives the anomalous gauge
variations on the boundary branes given in (\ref{gaugt4}). We then
check $\Gamma_{eff}$ and show how the gauge invariance is
achieved in (\ref{def8}).
$\Gamma_{eff}=\Gamma_{eff}(A_L,A_R,W)$ is the action
quantum corrected by boundary fermions $\psi_{L,R}$ and
is given at the functional level as
\bea
e^{i \Gamma_{eff}(A_L,A_R, W)}=\int D\psi_L D{\bar \psi}_L
D\psi_R D{\bar \psi}_R ~e^{i \Gamma_\psi(\psi_L,\psi_R,W,A_L,A_R)},
\label{effact}
\eea
where
\bea
\Gamma_\psi = \int d^4x ~[{\bar \psi}_L i \gamma^\mu
(\partial_\mu -i A_{L\mu}) \psi_L
+{\bar \psi}_R i \gamma^\mu(\partial_\mu -i A_{R\mu}) \psi_R 
-m ({\bar \psi}_L W \psi_R +h.c.)].
\label{def9}
\eea
$m$ can be complex in general. We have taken $m$ as real for simplicity.
$W$ is the Wilson line in (\ref{WilL}). Notice that a non-local
term ${\bar \psi}_L W \psi_R$ is included in the Lagrangian.
Under gauge transformation given in (\ref{gaugtr2}), (\ref{gaugtr2a}),
(\ref{gaugt3}) and (\ref{gaugt3d})
${\cal L}_\psi$ is gauge invariant but the functional measure
is not as shown in (\ref{gaugt6}).
For an infinitesimal transformation (\ref{gaugt3b}) and (\ref{gaugt3c})
we get
\bea
\delta \Gamma_{eff}&&=\Gamma_{eff}(A'_L,A'_R, W')
-\Gamma_{eff}(A_L,A_R, W) \nnb \\
&&=- \int d^4x ~[\epsilon^a_L G^a_L(A_L) +\epsilon^a_R G^a_R(A_R)] \nnb \\
&&=-\frac{1}{24\pi^2} \int d^4x ~\omega_4^1(A_L,\epsilon_L)
+\frac{1}{24\pi^2} \int d^4x ~\omega_4^1(A_R,\epsilon_R).
\label{gaugt7}
\eea
Putting this into (\ref{def8}) and using (\ref{gaugt4})
one sees that
\bea
\Gamma(A'_M,\psi'_L,\psi'_R)=\Gamma(A_M,\psi_L,\psi_R). \nnb
\eea
The anomalous variations cancel in the action under consideration
and the theory is gauge invariant.
\\

Now we try to show that $\Gamma_{eff}$ contains the WZW action
as a part and the WZW action given in the last subsection appears
in the effective theory when decoupling the heavy fermions
localized on the boundary branes as $m \to \infty$.
We change the variables $\psi_{L,R}$ to
$\chi_{L,R}$ and diagonalize the fermion mass matrix
using finite transformation $g_{L,R}$
\bea
&\chi_{L,R}=g_{L,R}\psi_{L,R}, ~~~W_\chi=g_L W g_R^{-1}=1,\nnb \\
& {\cal L}_\chi={\bar \chi}_L i \gamma^\mu (\partial_\mu
-i A_{L\mu}^{g_L}) \chi_L -m {\bar \chi}_L \chi_R
+(L \leftrightarrow R).
\label{gaugt8}
\eea
Computing the Jacobian when changing $\psi$ to $\chi$ 
with $\psi=g^{-1} \chi$ one obtains
\bea
 \Gamma_{eff}(A_L,A_R, \Sigma)=  \Gamma_\chi(A_L^{g_L},A_R^{g_R})
+ \Gamma_L(A_L^{g_L}, g_L)+ \Gamma_R(A_R^{g_R}, g_R),
\label{act1}
\eea
where $\Gamma_\chi$ is the effective action computed
in the base of $\chi$, and
\bea
\Gamma_L(A_L^{g_L},g_L(s))= \Gamma_A(A_L^{g_L},g_L(s)) , ~~
\Gamma_R(A_R^{g_R},g_R(s))= -\Gamma_A(A_R^{g_R},g_R(s)),
\label{act2}
\eea
where
\bea
\Gamma_A(A^g,g(s))&&= \frac{i}{24\pi^2}\int_{D_5} dsd^4x ~ 
~\omega_4^1(A^{g(s)}, g(s)\partial_s g^{-1}(s)),
\label{phase1} \\
A^g_\mu &&=g(x) A_\mu g^{-1}(x) +i g(x) \partial_\mu g^{-1}(x).
\label{gaugt9}
\eea
$D_5={\cal M}_4\times [0,1]$ is an auxiliary extention of
${\cal M}_4$.
(\ref{act2}) is the opposite of the Jacobian
changing variable $\chi$ to $\psi$ using $\chi=g \psi$.
$g_{L,R}(s)$($s \in [0,1]$) interpolates $g_{L,R}(s=0)=1$ and
$g_{L,R}(s=1)=g_{L,R}$. A computation of (\ref{phase1}) is
done in Appendix A and an explicit formula of $\Gamma_A$ is given.
\\

We can check the transformation properties of the effective
action as follows. There is a transformation leaving
${\cal L}_\chi$ invariant, {\it i.e.}
\bea
\chi \to V \chi, ~~ A^{g_L}_L \to (A^{g_L}_L)^V=A^{V g_L}_L,
~~ A^{g_R}_L \to (A^{g_L}_R)^V=A^{V g_R}_R.
\label{gaugt10}
\eea
This is a vector-like transformation.
The left-right transformation $W'= U_L W R^{-1}$ is manifested
on $g_L$ and $g_R$ as
\bea
g'_L = g_L U_L^{-1} ~~~ g'_R=g_R U_R^{-1}, \label{gaugt11}
\eea
which leaves $g'_L W' g'^{-1}_R=1$ invariant.
We note that the left-right transformation has not
effect on $A_L^{g_L}$ and $A_R^{g_R}$. It can be checked
that under the left-right transformation
\bea
(A'_L)^{g'_L}=(A_L^{U_L})^{g'_L}=A_L^{g'_L U_L}=A_L^{g_L}, \nnb 
\eea
same for $A_R^{g_R}$. That means ${\cal L}_\chi$ is not affected
by the left-right transformation.

In total $g_{L,R}$ can be transformed by $V$ and $U_{L,R}$, that is
\bea
g_{L,R} \to g'_{L,R}=V g_{L,R} U_{L,R}^{-1}.
\label{gaugt12}
\eea

Look at $\Gamma_L$ for example. What we need to compute
for $g'_L$ and $A'_L$ is
\bea
\Gamma_L((A^{U_L}_L)^{g'_L}, g'_L)
=\frac{i}{24\pi^2} \int_{D_5} ds d^4x ~\omega_4^1((A^{U_L}_L)^{g'_L(s)}, 
g'_L(s) \partial_s g'^{-1}_L(s)), \nnb
\eea
where $g'_L(s)$ interpolates $g'_L(s=0)=1$ and $g'_L(s=1)=g'_L$.
The integration is over gauge configurations from $A^{U_L}_L$
to $(A^{U_L}_L)^{g'_L}=A^{g'_L U_L}_L=A^{V g_L}_L$ in which
(\ref{gaugt12}) is used. Compared to
$\Gamma(A^{g_L}_L, g_L)$, the difference is in the integration
from gauge configurations $A_L$ to $A^{U_L}_L$ and $A^{g_L}_L$
to $A^{V g_L}_L$. For infinitesimal transformation
(\ref{gaugt3b}), (\ref{gaugt3c}) and $V=e^{i\epsilon_V}$ we have
\bea
\delta \Gamma_L &&= \Gamma_L( (A^{U_L}_L)^{g'_L},g'_L)-
\Gamma_L(A^{g_L}, g_L) \nnb \\
&&=  \frac{1}{24 \pi^2} \int d^4x ~\omega_4^1(A_L^{g_L},\epsilon_V)
 - \frac{1}{24\pi^2} \int d^4x ~\omega_4^1(A_L,\epsilon_L),
\label{gaugt13}
\eea
Similarly we can get transformation properties for $\Gamma_R$
with a difference on the sign.
We see that (\ref{gaugt13}) produces exactly what we expect
in (\ref{gaugt7}) for the left-right transformation. The extra
$V$ transformation corresponds to the freedom to reparametrize
$g_L$ and $g_R$, that is $W=(Vg_L)^{-1} V g_R$. 
The total effective action must be the same for different
parametrization, that is $\Gamma_{eff}$ is invariant under $V$
transformation.

So $\Gamma_\chi$ has to give the right transformation under
$V$, {\it i.e.}
\bea
\delta \Gamma_\chi(A^{g_L}_L,A^{g_R}_R)=-
\frac{1}{24 \pi^2} \int d^4x ~\omega_4^1(A_L^{g_L},\epsilon_V)
+\frac{1}{24 \pi^2} \int d^4x ~\omega_4^1(A_R^{g_R},\epsilon_V),\nnb
\eea
and this cancels variation in $\Gamma_L$ and $\Gamma_R$.
A conterterm, $\Gamma_B$, is known to give
the correct transformation properties~\cite{mm}, that is
\bea
\Gamma_B(B_L,B_R)&&=\frac{1}{48\pi^2} \int d^4x ~\varepsilon^{\mu\nu\rho\sigma}
~Tr[\frac{1}{2}(F^{BL}_{\mu\nu}+F^{BR}_{\mu\nu})(B_{R\rho}B_{L\sigma}
-B_{L\rho}B_{R\sigma}) \nnb \\
&&+i (B_{R\mu}B_{R\nu}B_{R\rho}B_{L\sigma}-
B_{L\mu}B_{L\nu}B_{L\rho}B_{R\sigma})-\frac{i}{2}
B_{R\mu}B_{L\nu}B_{R\rho}B_{L\sigma}], 
\label{Bard} 
\eea
where
\bea
F^{BL}_{\mu\nu}= \partial_\mu B_{L\nu}-\partial_\nu B_{L\mu}
-i [B_{L\mu},B_{L\nu}], ~~
F^{BR}_{\mu\nu}= \partial_\mu B_{R\nu}-\partial_\nu B_{R\mu}
-i [B_{R\mu},B_{R\nu}]. 
\label{def10}
\eea
$\Gamma_\chi$  must be obtained as~\footnote{
To author's knowledge $\Gamma_B$ known as the Bardeen's conterterm
is not yet obtained in computation of the effective action when
decoupling heavy chiral fermions in 4D chiral gauge theories.
The detailed computation is out of the scope of this paper
and a computation will be presented 
in a further publication.}
\bea
\Gamma_\chi(A_L^{g_L},A_R^{g_R})=\Gamma_B(A_L^{g_L},A_R^{g_R})
+ \textrm{polynomials invariant under $V$ },
\label{act3}
\eea

Armed with this observation we can write the anomalous
effective action $\Gamma_{WZW}$ as
\bea
\Gamma_{WZW}(A_L, A_R, W) 
=\frac{1}{2}[ \Gamma_B(A^{W^{-1}}_L,A_R)
+\Gamma_B(A_L,A^{W}_R)+\Gamma_L(A^{W^{-1}}_L,
W^{-1})+\Gamma_R(A^W_R,W) ].
\label{WZW2}
\eea
This is obtained as the mean of the actions in two cases
of $g_L=W^{-1}, g_R=1$ and $g_L=1, g_R=W$.
(\ref{WZW2}) reproduces (\ref{WZW1}) when changing
the integration over $s$ to integration over $y$ and taking
$g_L(s)=W^{-1}_y(s\times \pi R)$, $g_R(s)=1$ and
$g_L(s)=1$, $g_R(s)=W_y(s\times \pi R)$ in two cases.

\section{WZW action and anomalous chiral symmetry}\label{sec4}
In this section we construct WZW action
which is gauge invariant but gives anomalous chiral
transformation. This happens in the case chiral symmetry is not gauged.
Consider a toy model on $M_4\times I$ with coordinates
$x^M=(x^\mu,x^4=y)$. 
The space-like extra dimension $I$ is compact and
flat. It can be a cycle $I=S^1$ or orbifold
like $I=S^1/Z_2$, etc. Gauge fields $B_M(x^\mu,y)$
of gauge group $G$ propagate in the bulk.
Depending on the boundary condition, the gauge group on 
the brane at $y=0$, $H$ which is $G$ restricted at $y=0$, 
can be smaller or equal to the gauge group in the bulk:
$H \subseteq G$.
We explain in detail the toy model as follows.

We define the periodicity of the compact space $I$
using the equivalence class
\bea
y+2\pi R \sim  y.
\eea
Gauge fields are equivalent up to gauge transformation:
\bea
B_M(x^\mu,y+2 \pi R)= \Omega B_M(x^\mu,y)\Omega^{-1} 
+i \Omega \partial_M \Omega^{-1}.
\label{pred}
\eea
Up to a global phase $\Omega=\Omega(x^\mu,y)$ is an element of $G$.
Under a local gauge transformation $U(x^\mu,y)$, $B_M$ and $\Omega$ are
transformed according to
\bea
B'_M(x^\mu,y)&&= U(x^\mu,y) B_M(x^\mu,y)U^{-1}(x^\mu,y) 
+i U(x^\mu,y) \partial_M U^{-1}(x^\mu,y);
\label{gautr1}\\
\Omega'(x^\mu, y)&&= U(x^\mu, y+2\pi R) \Omega(x^\mu,y) U^\dagger(x^\mu,y).
\label{gautr2}
\eea

We define a Wilson loop $W(x^\mu)$ as
\bea
W(x^\mu)= {\cal P} e^{i\int^{2 \pi R}_0 dy ~B_4(x^\mu,y)}  
\times \Omega(x^\mu, 0),
\label{WilLP}
\eea
where ${\cal P}$ is the path-ordering operator.
Using (\ref{gautr1}) and (\ref{gautr2}), 
$W(x^\mu)$ can be shown to be gauge covariant:
\bea
W'&&= P e^{i \int^{2 \pi R}_0 dy ~B'_4} \times \Omega'(x^\mu, 0)\nnb \\
  &&= U(x^\mu, 0) ~{\cal P} e^{i \int^{2 \pi R}_0 dy ~B_4} 
~U^\dagger(x^\mu, 2 \pi R)
      \times \Omega'(x^\mu, 0) \nnb \\
  &&= U(x^\mu,0) ~W ~U^\dagger(x^\mu,0).
\label{gautr3}
\eea

We introduce the gauge fields $V_\mu$ and $A_\mu$ on the brane at $y=0$ as
\bea
V_\mu(x^\mu)=B_\mu(x^\mu,y=0),~~A_\mu=0.
\label{def11}
\eea
For convenience of later discussion we introduce $A_L$ and $A_R$
\bea
A_R=V_\mu+A_\mu, ~~A_L=V_\mu-A_\mu,
\label{def11a}
\eea
so that $V_{\mu\nu}$ and $A_{\mu\nu}$ defined in (\ref{streng1}) 
and (\ref{streng2}) are
\bea
V_{\mu\nu}= \frac{1}{2}(F^R_{\mu\nu}+ F^L_{\mu\nu}),~~
A_{\mu\nu}= \frac{1}{2}(F^R_{\mu\nu}- F^L_{\mu\nu}).
\label{def11b}
\eea
$F^{L,R}_{\mu\nu}=\partial_\mu A_{L,R\nu}-\partial_\nu A_{L,R\mu}
-i [A_{L,R\mu},A_{L,R\nu}]$ are the field strengths for $A_{L,R}$.
On the brane at $y=0$ the gauge transformation $U(x^\mu,y)$ is
reduced to be $U(x^\mu)$, {\it i.e.}
\bea
&U(x^\mu)=U(x^\mu,y=0), ~~ W'= U(x^\mu)W U^{-1}(x^\mu),
\label{gaugtr4a} \\
&A'_{L,R\mu}=U(x^\mu) A_{L,R\mu} U^{-1}(x^\mu)+i U(x^\mu)\partial_\mu 
U^{-1}(x^\mu).
\label{gautr4}
\eea
\\

We introduce $W(s)$ for $s \in [0,1]$ with condition
\bea
W(s=0) =1, ~~~ W(s=1)=W.
\label{def11d}
\eea
$W(s)$ interpolates unit matrix and $W$.
We can construct the WZW action as 
\bea
\Gamma_{WZW}=\frac{i}{8\pi^2} \int^1_0 ds \int_{{\cal M}_4} d^4x
~\omega^2_4(V^{W(s)},A^{W(s)},W(s)),
\label{act4}
\eea
where
\bea
\omega^2_4(V^{W(s)},A^{W(s)},W(s))
&&= \varepsilon^{\mu\nu\rho\sigma} Tr[ W^{-1}(s)\frac{\partial W(s)}
{\partial s} (\frac{1}{4} V^{W(s)}_{\mu\nu}V^{W(s)}_{\rho\sigma}
+\frac{1}{12}A^{W(s)}_{\mu\nu}A^{W(s)}_{\rho\sigma} \nnb \\
&&+\frac{2i}{3}(V^{W(s)}_{\mu\nu} A^{W(s)}_\rho A^{W(s)}_\sigma
+A^{W(s)}_\mu V^{W(s)}_{\nu\rho} A^{W(s)}_\sigma
+A^{W(s)}_\mu A^{W(s)}_\nu V^{W(s)}_{\rho\sigma}) \nnb \\
&&-\frac{8}{3} A^{W(s)}_\mu A^{W(s)}_\nu A^{W(s)}_\rho A^{W(s)}_\sigma],
\label{def12}
\eea
and
\bea
V^{W(s)}_{\mu\nu}
&&=\frac{1}{2} [F^R_{\mu\nu}+W^{-1}(s)F^L_{\mu\nu}W(s) ],~~
A^{W(s)}_{\mu\nu}
=\frac{1}{2}[ F^R_{\mu\nu}-W^{-1}(s)F^L_{\mu\nu}W(s) ],
\label{def12a} \\
A^{W(s)}_\mu
&&=\frac{1}{2}[ A_{R\mu}-W^{-1}(s)A_{L\mu} W(s)
-i W^{-1}(s) \partial_\mu W(s) ].
\label{def12b}
\eea

It's easy to check the transformation property of (\ref{act4}).
Under the gauge transformation given in (\ref{gaugtr4a})
and (\ref{gautr4}) we have
\bea
&W'(s)=U(x^\mu)W(s) U^{-1}(x^\mu),~~
A'^{W'(s)}_\mu= U(x^\mu) A^{W(s)}_\mu U^{-1}(x^\mu), 
\\
&V'^{W'(s)}_{\mu\nu}= U(x^\mu) V^{W(s)}_{\mu\nu} U^{-1}(x^\mu),~~
A'^{W'(s)}_{\mu\nu}= U(x^\mu) A^{W(s)}_{\mu\nu} U^{-1}(x^\mu).
\eea
$W'(s)$ satisfies the condition (\ref{def11d}) as can be easily
seen. So it's clear that (\ref{act4}) and (\ref{def12}) are invariant
under the vector-like gauge transformation.

Consider next the chiral transformation
\bea
&W'= U_A^{-1} W U_A^{-1}, 
\label{gautr5} \\
&A'_{L\mu}= U_A^{-1} A_{L\mu} U_A +i U_A^{-1} \partial_\mu U_A,~~
A'_{R\mu}= U_A A_{R\mu} U_A^{-1} +i U_A \partial_\mu U_A^{-1}.
\label{gautr6}
\eea
$W'(s)$, which interpolates the unit matrix and $W'=U_A^{-1} W U_A^{-1}$,
is then introduced into (\ref{act4}).
If we write
\bea
W'(s)=U^{-1}_A(x^\mu) {\wti W}(s) U^{-1}_A(x^\mu),
\label{def13}
\eea
we then have ${\wti W}(s)$ which interpolates ${\wti W}(s=0)=U^2_A(x^\mu)$
and ${\wti W}(s=0)=W$. One can check
\bea
&A'^{W'(s)}_\mu= U_A(x^\mu) A^{{\wti W}(s)}_\mu U^{-1}_A(x^\mu) \nnb \\
&V'^{W'(s)}_{\mu\nu}= U_A(x^\mu) V^{{\wti W}(s)}_{\mu\nu} U^{-1}_A(x^\mu),~~
A'^{W'(s)}_{\mu\nu}= U_A(x^\mu) A^{{\wti W}(s)}_{\mu\nu} U^{-1}_A(x^\mu), \nnb
\eea
and the chirally transformed $\omega_4^2$ can be rewritten as
\bea
\omega^2_4(V'^{W'(s)},A'^{W'(s)},W'(s))=
\omega^2_4(V^{{\wti W}(s)},A^{{\wti W}(s)},{\wti W}(s)).
\label{gautr7}
\eea
Plugging it into (\ref{act4}) and comparing it with the un-transformed case
the difference is the integration
from $W(s)=1$ to $W(s)=U^2_A(x^\mu)$.
For an infinitesimal transformation $U_A=e^{i \epsilon_A}$ we
have
\bea
\delta \Gamma_{WZW}=- \epsilon^a_A G^a_A(V,A),
\label{gautr8}
\eea
where $\epsilon_A=\sum_a T^a \epsilon^a_A$.
This is exactly the anomaly in VA form.
\\

The action (\ref{act4}) can also be obtained as arising from
a decoupling fermion. We introduce fermions 
$\psi=(\psi_L,\psi_R)$ which are localized on the brane at $y=0$. 
They are transformed under the gauge transformation (\ref{gautr4}) as
\bea
\psi'_{L,R}(x^\mu)= U(x^\mu) \psi_{L,R}(x^\mu).
\label{gautr4a}
\eea
Consider the effective action $\Gamma_{eff}$
\bea
e^{i\Gamma_{eff}(A_L,A_R,W)}= \int D\psi_L D{\bar \psi}_L
 D\psi_R D{\bar \psi}_R~ e^{i\Gamma_\psi}.
\label{def14}
\eea
$A_{L,R}$ and $W$ as introduced in (\ref{WilLP}) and
(\ref{def11a}) are considered as external fields and
\bea
\Gamma_\psi &&=
\int d^4x ~[\bar{\psi} i \gamma^\mu(\partial_\mu
-i T^a V^a_\mu )\psi -m ({\bar \psi}_L W \psi_R + h.c.) ], 
\nnb \\
&& =\int d^4x ~[\bar{\psi}_L i \gamma^\mu(\partial_\mu
-i A_{L\mu}) \psi_L +\bar{\psi}_R i \gamma^\mu(\partial_\mu
-i A_{R\mu})\psi_R-m ({\bar \psi}_L W \psi_R + h.c.) ].
\label{act5}
\eea
In general $m$ is complex. We have chosen $m$ to be real
for convenience.
Action (\ref{act5}) is gauge invariant
under the gauge transformation give in (\ref{gaugtr4a}), (\ref{gautr4})
and (\ref{gautr4a})

Gauge couplings in (\ref{act5}) with gauge fields introduced
in (\ref{def11}) and (\ref{def11a}) take the form in which
only the vector part is gauged. As discussed before, the
anomaly takes the VA form for which the anomaly is completely
shifted to the axial-vector current and the vector current
is covariantly conserved. Anomaly of VA form can also be
obtained in the functional level. It is achieved by carefully
treating the Dirac operator in the
evaluation of the Jacobian~\cite{bert,bmnt}. 
Consequently, under
a chiral transformation (\ref{gautr5}) and (\ref{gautr6})
supplemented by
\bea
\psi'_R=U_A \psi_R, ~~ \psi'_L=U_A^{-1} \psi_L,
\label{gautr5a}
\eea
$\Gamma_{eff}$ changes
by an anomalous term due to changing the measure of
$\psi'$ to that of $\psi$. For an infinitesimal transformation
$U_A=e^{i \epsilon_A}$ it gives
\bea
\delta \Gamma_{eff} && =\Gamma_{eff}(A'_L,A'_R,W')
-\Gamma_{eff}(A_L,A_R,W)= -\epsilon^a_A G^a_A(V,A),
\label{gautr9}
\eea
where $G^a_A(V,A)$ is given in (\ref{ano3}) and
$V_\mu$ and $A_\mu$ shown in (\ref{def11a}).

Similar to the case in the last section we can obtain
the anomalous action by doing a finite chiral transformation
\bea
\psi_R=\xi^{-1} \chi_R, ~~\psi_L=\xi \chi_L, ~~\textrm{ with}
~W=\xi^2.
\label{gautr10}
\eea
We introduce $\xi(s)$ and $W(s)=\xi^2(s)$with condition
\bea
&\xi(s=0)=1, ~~\xi(s=1)=\xi, \nnb \\
&W(s=0)=1, ~~W(s=1)=W. \nnb
\eea
WZW action is obtained as a continuous integration of chiral transformation
\bea
\Gamma_{WZW}=i \int^1_0 ~ds \int_{{\cal M}_4} d^4x~Tr\{
\frac{1}{2}[\xi(s)\partial_s\xi^{-1}(s)-\xi^{-1}(s)\partial_s \xi(s)]
G^a_A(V^{\xi(s)},A^{\xi(s)})\},
\label{act6}
\eea
It is the opposite of the phase integration of changing $\chi_R=\xi \psi_R$
to $\psi_R$. $A^{\xi(s)}_\mu$ in (\ref{act6}) is the axial-vector part of
$A^{\xi^{-1}(s)}_{L\mu}$ and $A^{\xi(s)}_{R\mu}$ computed
using (\ref{gaugt9}), that is
\bea
A^{\xi(s)}_\mu &&= \frac{1}{2}[
\xi(s) A_{R\mu} \xi^{-1}(s) + i\xi(s)\partial_\mu \xi^{-1}(s)
-\xi^{-1}(s) A_{L\mu} \xi(s)- i\xi^{-1}(s) \partial_\mu\xi(s)] \nnb \\
&&= \frac{1}{2} \xi(s) [A_{R\mu}- W^{-1}(s) A_{L\mu} W(s)-i W^{-1}(s)
\partial_\mu W(s)] \xi^{-1}(s) \nnb \\
&& =\xi(s) A^{W(s)}_\mu \xi^{-1}(s).
\label{gautr11}
\eea
Similarly we have
\bea
&&V^{\xi(s)}_{\mu\nu}=\xi(s) V^{W(s)}_{\mu\nu} \xi^{-1}(s),~~
A^{\xi(s)}_{\mu\nu}=\xi(s) A^{W(s)}_{\mu\nu} \xi^{-1}(s),
\label{gautr12}
\eea
Noticing further that
\bea
\xi(s)\partial_s\xi^{-1}(s)-\xi^{-1}(s)\partial_s \xi(s)
=-\xi(s) W^{-1}(s) [\partial_s W(s)] \xi^{-1}(s), \nnb
\eea
it is then easy to rewrite $\Gamma_{WZW}$ in (\ref{act6}) to be the
form in (\ref{act4}).

\section{Summary}\label{sec5}
Using examples in 3D and 5D we have shown that Wilson line or
Wilson loop along the compact space-like dimension can be used
to construct the WZW action. If $W$ is a Wilson line
which links the two boundary branes, $W$ transforms
as bifundamental under the gauge groups of the two boundaries.
Gauge symmetries at two branes can be considered in 4D point
of view as two independent symmetries though 
they are parts of the 5D gauge symmetry.
Anomalous action constructed using $W$ gives
anomalous gauge variations (of LR form) 
on two boundary branes. On the other hand, the Wilson loop
transforms under a single gauge symmetry of one brane.
The anomalous action constructed using Wilson loop will then be 
invariant under vector-like gauge transformation.
The WZW action gives the anomalous variations under the
chiral transformation. 

The WZW action constructed using
Wilson line gives non-local interactions for gauge fields
on separated boundary branes via the link field, {\it i.e.} the
Wilson line. It should be
understood in odd dimensional bulk, rather than in the even-dimensional
boundary branes. We have used this action to generalize the gauge
theory in 3D and 5D space-times with boundaries. 
The generalized gauge theory
includes actions of pure Yang-Mills, Chern-Simons and WZW
in the bulk. The gauge invariance of the 
theory is achieved by requiring the
that the anomalous gauge variations of Chern-Simons and WZW actions
cancel on the boundary space-times.
We expect that this procedure can be generalized to odd dimensions
larger than five. The action constructed is expected to be 
of the form of WZW action in even dimensions higher than 
four~\cite{wu}. We have constructed action in flat spce-time
background. We expect procedure descriped in this paper
can be generalized to curved space-time background.
The generalization to include gravity is a problem
and is worth further study.

We tried to show that WZW constructed using Wilson line or
Wilson loop can be understood as arising from
integrating out heavy fermion. In 5D examples, these fermions are
localized on 4D boundaries. They can not couple to 
extra dimensional gauge field $A_4$ using ${\bar \psi} \gamma_5 \psi A_4
\psi$ since there is no corresponding kinetic term 
${\bar \psi}\gamma_4 \partial_4\psi$ on the boundary brane.
The coupling using Wilson line or Wilson loop, such as
\bea
m({\bar \psi}_L W \psi_R +h.c.), \nnb
\eea
is perfectly allowed. This is a natural way for
$A_4$ degrees of freedom to couple to localized fermion.
Integrating out localized fermion as sending $m \to \infty$
results in the WZW actions which
manifest the gauge or global anomalies of fermions
in the effective theory. Ref. ~\cite{Hill1} approached
the point in the case of using Wilson line.
The approach taken in this paper is easier to
see how it happens.
We note that the procedure shown in this paper is
not a proof. To have a proof one needs to work in
the quantized theory in odd dimensional space-times
with boundaries. This is out of the scope of this
paper.

The WZW action constructed using Wilson line or
Wilson loop has interesting applications and
phenomelogical implications. One application of the action is
to use it to implement the anomaly inflow mechanism \cite{liao}
with chiral fermions localized on the boundary space-times. 
It is similar to the setup of the quantum Hall effect,
although we have two boundaries rather than the single
boundary in quantum Hall effect.
Similar in the sigma model, one can write the Wilson line 
or the Wilson loop as $W=e^{2i {\tilde \pi}/F}$ where
$F=1/(2\pi R)$ or $F=1/(\pi R)$. 
The gauge field $A_4$ is then like meson fields.
We note that KK modes that are not periodic around
the extra dimension can contribute to the Wilson
loop, {\it e.g.} the anti-periodic modes in some
models. Then the meson fields ${\tilde \pi}$ as defined
in $W=e^{2 i {\tilde \pi}/F}$ can contain massive
KK modes. In this case processes like ${\tilde \pi}
\to 2 \gamma$ is allowed if the gauge symmetry on
the boundary brane includes that of electromagnetism.
To understand the phenomenology of the WZW action in 5D,
its KK mode expansion needs to be
studied.

%%%%%%%%%%%%%%%%%%%%%%%%%%%%%%%%%%%%%%%%%%%%%%%%%%%%%%%%%

\section*{\bf Appendix A. Evaluation of phase integration}\nonumber

The phase integration can be conveniently computed using
differential forms. Consider the integration on $D_5={\cal M}_4 \times [0,1]$
\bea
\Gamma_A(A^g,g(s))&&= \frac{i}{24\pi^2}\int_{D_5} dsd^4x ~ Tr[g(s)
\frac{\partial g^{-1}(s)}{\partial s} 
\varepsilon^{\mu \nu \rho \sigma}
\partial_\mu (A_{\nu}^{g(s)} \partial_\rho A^{g(s)}_{\sigma}
-\frac{i}{2} A^{g(s)}_{\nu}A^{g(s)}_{\rho}A^{g(s)}_{\sigma})].~
\label{A01}
\eea
Using
\bea
A_{g(s)}=A^{g(s)}_\mu dx^\mu, ~~~
F_{g(s)}=\frac{1}{2} F^{g(s)}_{\mu \nu} dx^\mu d^\nu
=dA_{g(s)} -i A_{g(s)}^2, ~~~ d^2 =0,
\label{A02}
\eea
one can write (\ref{A01}) as
\bea
\Gamma_A(A^g,g(s))
&& =\frac{i}{24\pi^2}\int_{D_5} ds ~Tr[g(s) 
\frac{\partial g^{-1}(s)}{\partial s} d( A_{g(s)} d A_{g(s)}
-\frac{i}{2} A_{g(s)} A_{g(s)} A_{g(s)}) ] \nnb \\
&& =\frac{i}{24\pi^2}\int^1_0 ds ~Tr[g(s)
\frac{\partial g^{-1}(s)}{\partial s}( F_{g(s)}^2 \nnb \\
&& +\frac{i}{2}(A_{g(s)}^2 F_{g(s)}+A_{g(s)}F_{g(s)}A_{g(s)}
+F_{g(s)} A_{g(s)}^2) -\frac{1}{2}A_{g(s)}^4) ].
\label{A03}
\eea

We extend the gauge fields $A$ to ${\cal A}$ with a fictitious dimension
and fictitious gauge field ${\cal A}_s$
\bea
&{\cal A}_\mu =A_\mu,~~ {\cal F}_{\mu \nu}=F_{\mu \nu},
~~{\cal A}_s=0, \label{A04} \\
&{\cal A}^{g(s)}_s
=g(s) {\cal A}_s g(s)^{-1} +i g(s) \partial_s g(s)^{-1}
=i g(s) \partial_s g(s)^{-1}, \label{A05} \\
&{\cal F}_{\mu s}=\partial_\mu {\cal A}_s-\partial_s {\cal A}_\mu 
-i [ {\cal A}_\mu, {\cal A}_s]=0, \label{A06} \\
&{\cal F}_{\mu s}^{g(s)}= g(s) {\cal F}_{\mu s} g(s)^{-1} =0.
\label{A07}
\eea
We can extend the 4-forms in (\ref{A03}) to 5-forms
with $\varepsilon^{s0123}=\varepsilon^{0123}=-1$
($\varepsilon_{s0123}=-\varepsilon_{0123}=-1$) and get
\bea
\Gamma_A(A^g,g(s))=
\frac{1}{24\pi^2}~ \int_{D_5} ~\omega_5({\cal A}_{g(s)},{\cal F}_{g(s)}),
\label{A08}
\eea
where $\omega_5({\cal A}_{g(s)},{\cal F}_{g(s)})$ is the Chern-Simons
5-forms:
\bea
\omega_5({\cal A}_{g(s)},{\cal F}_{g(s)})=
Tr[{\cal A}_{g(s)} {\cal F}_{g(s)}^2
+\frac{i}{2} {\cal A}_{g(s)}^3 {\cal F}_{g(s)}-\frac{1}{10} {\cal A}_{g(s)}^5].
\label{A09}
\eea
Compared to the last term in (\ref{A03}), a factor $1/5$
arises in ${\cal A}_{g(s)}^5$ term due to the cyclic symmetry.

$\omega_5$ is not a gauge invariant. One can write~\cite{zwz}
\bea
\omega_5({\cal A}_{g(s)},{\cal F}_{g(s)})&&=
\omega_5({\cal A}+ P, {\cal F}) \nnb \\
&& = \omega_5({\cal A}, {\cal F})+ d \alpha_4({\cal A},P)
 +\omega_5(P,0),
\label{A10}
\eea
where $P=i (dg(s)^{-1}) g(s)$ is a 1-form and
\bea
\alpha_4({\cal A},P)=
-Tr[\frac{1}{2}P({\cal A} d{\cal A}+d{\cal A}{\cal A})
-\frac{i}{2} P {\cal A}^3 + \frac{i}{2}P^3 {\cal A}
+\frac{i}{4} P {\cal A} P {\cal A}].
\label{A11}
\eea
In this case $\omega_5({\cal A}, {\cal F})=0$ because of ${\cal A}_s=0$.
One can check (\ref{A10}) explicitly using $d P= - iP^2$.

Putting (\ref{A10}) into (\ref{A08}) and integrate $d \alpha_4$ term
over interval $[0,1]$ one obtain
\bea
i\Gamma_A(A^g,g(s))= \frac{i}{24\pi^2}~ \int_{D_5} ~\omega_5(P,0)
+\frac{i}{24\pi^2}~ \int_{{\cal M}_4}~
\alpha_4({\cal A}(s),P(s)) \bigg|^{s=1}_{s=0}.
\label{A12}
\eea
Since $P(s=0)=0$ if reduced to ${\cal M}_4$ we can get the action
\bea
\Gamma_A(A^g,g(s))&&= \frac{-i}{240\pi^2}~ \int_{D_5} ds ~d^4 x
~Tr[\epsilon^{PQRST} \frac{\partial g^{-1}}{\partial x^P} g
\frac{\partial g^{-1}}{\partial x^Q} g
\frac{\partial g^{-1}}{\partial x^R} g
\frac{\partial g^{-1}}{\partial x^S} g
\frac{\partial g^{-1}}{\partial x^T} g] \nnb \\
&&+\frac{-i}{48\pi^2}~\int_{{\cal M}_4} d^4 x~
\epsilon^{\mu\nu\rho\sigma} Tr[ \frac{\partial g^{-1}}{\partial x^\mu} g
(A_\nu \partial_\rho A_\sigma +\partial_\nu A_\rho A_\sigma
-iA_\nu A_\rho A_\sigma) \nnb \\
&&-i \frac{\partial g^{-1}}{\partial x^\mu} g
\frac{\partial g^{-1}}{\partial x^\nu} g
\frac{\partial g^{-1}}{\partial x^\rho} g A_\sigma
-\frac{1}{2} \frac{\partial g^{-1}}{\partial x^\mu} g
A_\nu \frac{\partial g^{-1}}{\partial x^\rho} g A_\sigma],
\label{A13}
\eea
where $P,Q,R,S,T$ run over $s,0,1,2,3$ and $\varepsilon^{s\mu\nu\rho\sigma}
=\varepsilon^{\mu\nu\rho\sigma}$ has been used.

\end{document}